\documentclass{nic-series}
\usepackage{textgreek} 

\begin{document} 

\title{Realising the LOFAR Two-Metre Sky Survey \\
{\normalsize - using the supercomputer JUWELS at the Forschungszentrum J\"ulich - } }

\author{A. Drabent\inst{1} \and  M. Hoeft\inst{1} \and A.~P. Mechev\inst{2} \and J.~B.~R. Oonk\inst{2,3,4} \and T.~W. Shimwell\inst{2,3} \and F. Sweijen\inst{2} \and  A. Danezi\inst{4}, C. Schrijvers\inst{4}, C. Manzano\inst{5}, O. Tsigenov\inst{5}, R.-J. Dettmar\inst{6} \and M. Br\"uggen\inst{7} \and D.~J. Schwarz\inst{8}}

\institute{Th\"uringer Landessternwarte,\\
         Sternwarte 5, D-07778 Tautenburg, Germany\\
         \email{\{alex, hoeft\}@tls-tautenburg.de}
          \and
          Leiden Observatory, Leiden University,\\
          P.O. Box 9513, NL-2300 RA Leiden, The Netherlands
          \and
          Netherlands Institute for Radio Astronomy (ASTRON),\\
          Oude Hoogeveensedijk 4, NL-7991 PD Dwingeloo, The Netherlands
          \and
          SURFsara,\\
          P.O. Box 94613, NL-1090 GP Amsterdam, The Netherlands
          \and
          J\"ulich Supercomputing Centre, Institute for Advanced Simulation (IAS),\\
          Forschungszentrum J\"ulich, D-52425 J\"ulich, Germany
          \and
          Ruhr-Universit\"at Bochum, Astronomisches Institut,\\
          Universit\"atsstra{\ss}e 150, D-44780 Bochum, Germany
          \and
          Hamburger Sternwarte,\\ 
          Gojenbergweg 112, D-21029 Hamburg, Germany
          \and
          Fakult\"at f\"ur Physik, Universit\"at Bielefeld,\\
          Universit\"atsstra\ss{}e 25, D-33615 Bielefeld, Germany}

\maketitle

\begin{abstracts}
The new generation of high-resolution broad-band radio telescopes, like the Low Frequency Array (LOFAR), produces, depending on the level of compression, between 1 to 10\,TB of data per hour after correlation. Such a large amount of scientific data demand powerful compu\-ting resources and efficient data handling strategies to be mastered. The LOFAR Two-metre Sky Survey (LoTSS) is a Key Science Project (KSP) of the LOFAR telescope. It aims to map the entire northern hemisphere at unprecedented sensitivity and resolution. The survey consist of 3\,168 pointings, requiring about 30\,PBytes of storage space. As a member of the German Long Wavelength Consortioum (GLOW) the Forschungszentrum J\"ulich (FSZ) stores in the Long Term Archive (LTA) about 50\,\% of all LoTSS observations conducted to date. In collaboration with SURFsara in Amsterdam we developed service tools that enables the KSP to process LOFAR data stored in the J\"ulich LTA  at the supercomputer JUWELS in an automated and robust fashion. Through our system more than 500 out of 800 existing LoTSS observations have already been processed with the {\tt prefactor} pipeline. This pipeline calibrates the direction-independent instrumental and ionospheric effects and furthermore reduces the data size significantly. For continuum imaging, this processing pipeline is the standard pipeline that is executed before more advanced processing and image reconstruction methods are applied.
\end{abstracts}

\section{Introduction}
The {\bf Low Frequency Array}\cite{LOFAR,2013A&A...556A...2V} (LOFAR) is a novel radio telescope. It has been developed, built, and is operated by the Netherlands Institute for Radio Astronomy (ASTRON). In addition, there is a strong contribution by institutes in other European countries. LOFAR consists of 38 stations in the Netherlands, and 13 stations outside the Netherlands, six of them in Germany, three in Poland, and one in each of France, UK, Sweden, and Ireland. The array is still expanding with a new station in Latvia and Italy\cite{LOFAR_Italy}. All LOFAR partners together form the International LOFAR Telescope (ILT). 

LOFAR is an interferometer observing in the poorly explored frequency range between 10 and 240\,MHz with unprecedented sensitivity and spatial resolution in comparison to preceding telescopes operating at the low-frequency regime. It therefore opens up a new window to the Universe.

\section{The LOFAR Two-Metre Sky Survey (LoTSS)}

\begin{figure}[htbp]
\centering
\includegraphics[width = \textwidth]{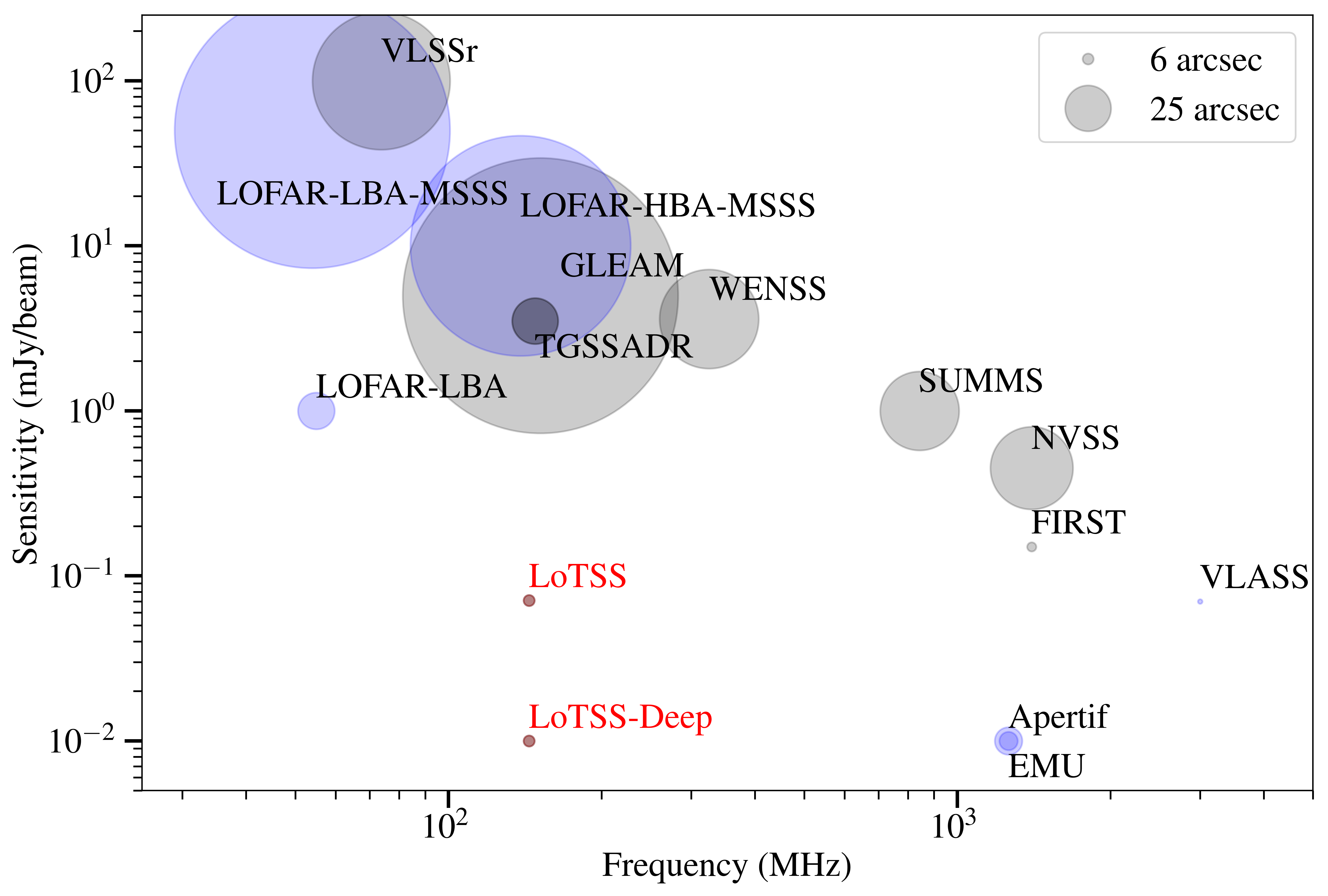}
\caption{Sensitivity versus frequency of selected recent (grey), forthcoming (blue), and LOFAR (red) radio sky surveys. The size of the dot resembles the square root of the resolution of the survey. Image from T.~W. Shimwell\cite{2019A&A...622A...1S}.}
\label{fig:LoTSS}
\end{figure}

For LOFAR several Keys Science Projects (KSPs) have been defined. These large projects demand a large fraction of the available observing time to be granted. The LOFAR Two-metre Sky Survey (LoTSS)\cite{LoTTS,2017A&A...598A.104S} is a project of the LOFAR Surveys KSP (SKSP) and is observing the entire northern sky. At optimal declinations LoTSS produces 6$^{\prime\prime}$ images with sensitivities below 100 \textmu Jy\,beam$^{-1}$ at a central frequency of 144\,MHz, using a bandwidth of 48\,MHz, see Fig.\,\ref{fig:LoTSS}.
To cover the entire northern sky 3\,168 observations will be carried out, each of which requires 8\,hours of observing time. This will eventually lead to an extraordinarily detailed map with 120 billion pixels and an expected amount of about 30\,PByte of data to be stored in the archives.

\begin{figure}[htbp]
\centering
\includegraphics[width = \textwidth]{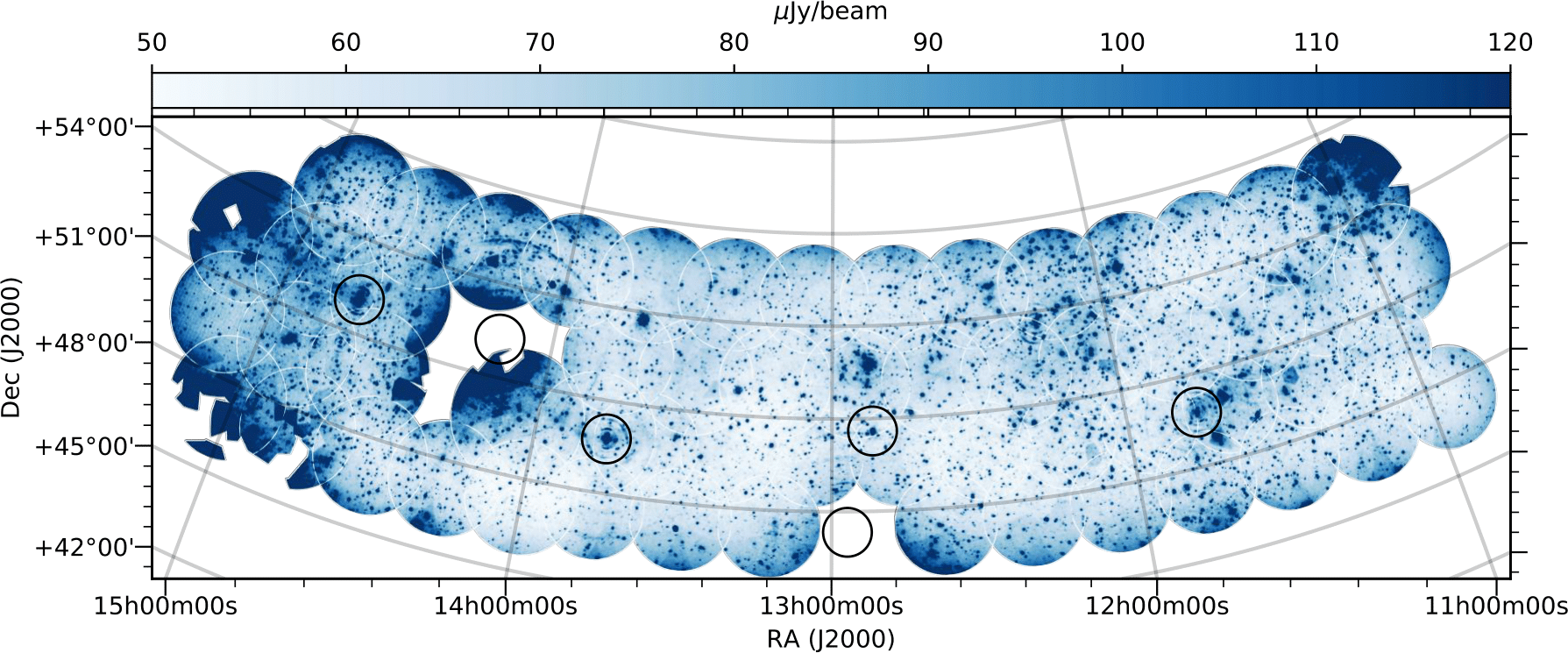}
\caption{Noise image of the HETDEX field published in the first LoTSS data release. The median noise level is 71\,\textmu Jy\,beam$^{-1}$. Black circles mark the locations of bright radio sources from the 3Cr catalogue and indicate problematic areas for the current calibration techniques. Image from Shimwell et al. (2019)\cite{2019A&A...622A...1S}.}
\label{fig:LoTSS_DR1}
\end{figure}

\begin{figure}[htbp]
\centering
\includegraphics[trim = 0 16 0 16, clip, width = 0.45\textwidth]{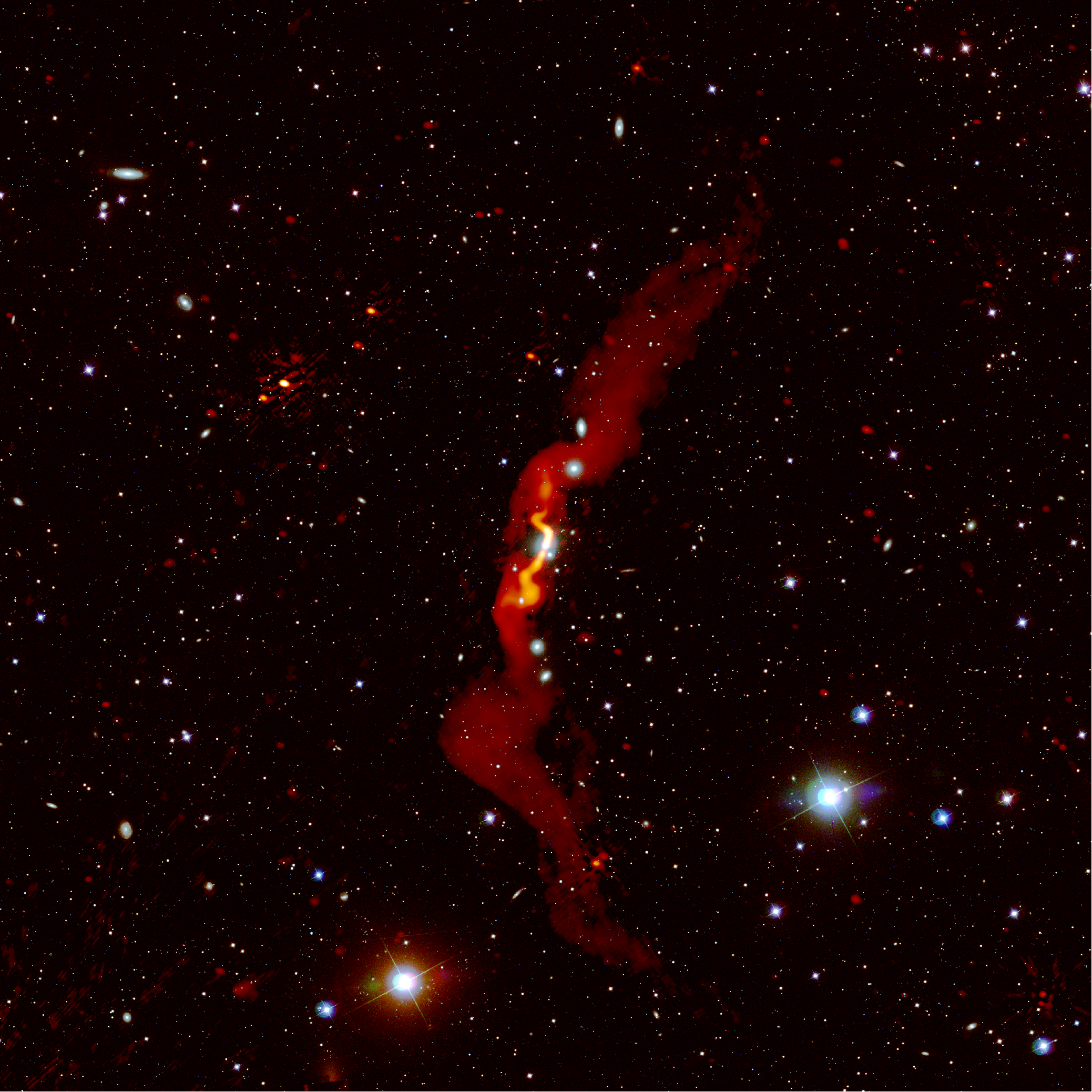}
\includegraphics[width = 0.45\textwidth]{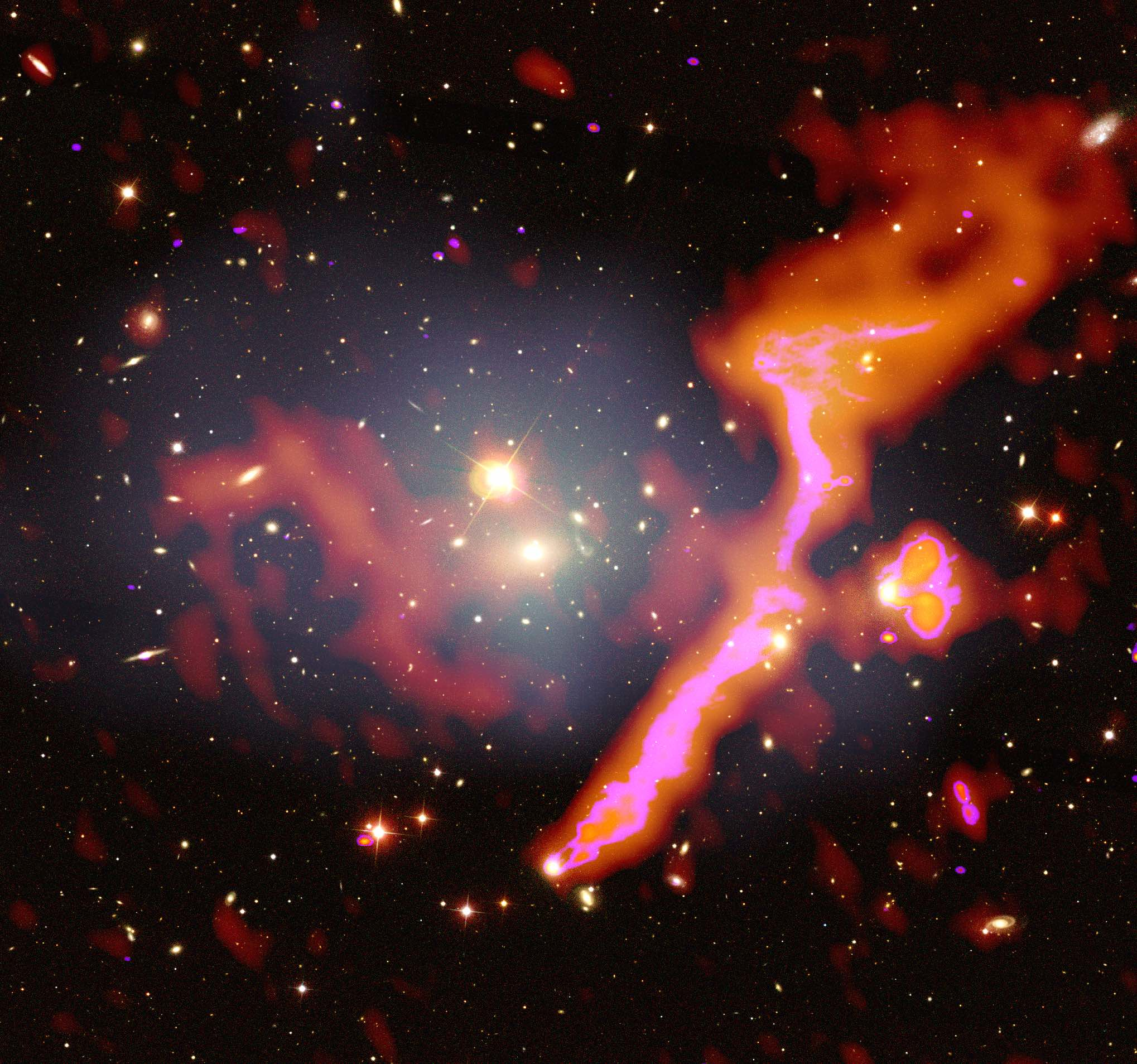}
\caption{Total intensity radio maps from the LoTSS (in red colour scale) overlaid on top of an optical image. Credit: LOFAR Surveys Team. \textbf{Left panel:} The radio galaxy 3C\,31. The observation reveals that the size of the galaxy is more than 3 million light years (Heesen et al., 2018)\cite{2018MNRAS.474.5049H}. \textbf{Right panel:} The galaxy cluster Abell\,1314. LOFAR reveals the presence of large-scale diffuse radio emission that traces highly-relativistic electrons in the intracluster medium embedded in a $\sim$\,\textmu G strong cluster-wide magnetic field. It is believed that such emission was caused by a merger with another cluster. Thermal bremsstrahlung emission detected with the Chandra X-ray observatory is shown in grey (Wilber et al., 2019)\cite{2019A&A...622A..25W}.}
\label{fig:LoTSS_results}
\end{figure}

\begin{figure}[htbp]
\centering
\includegraphics[width = \textwidth]{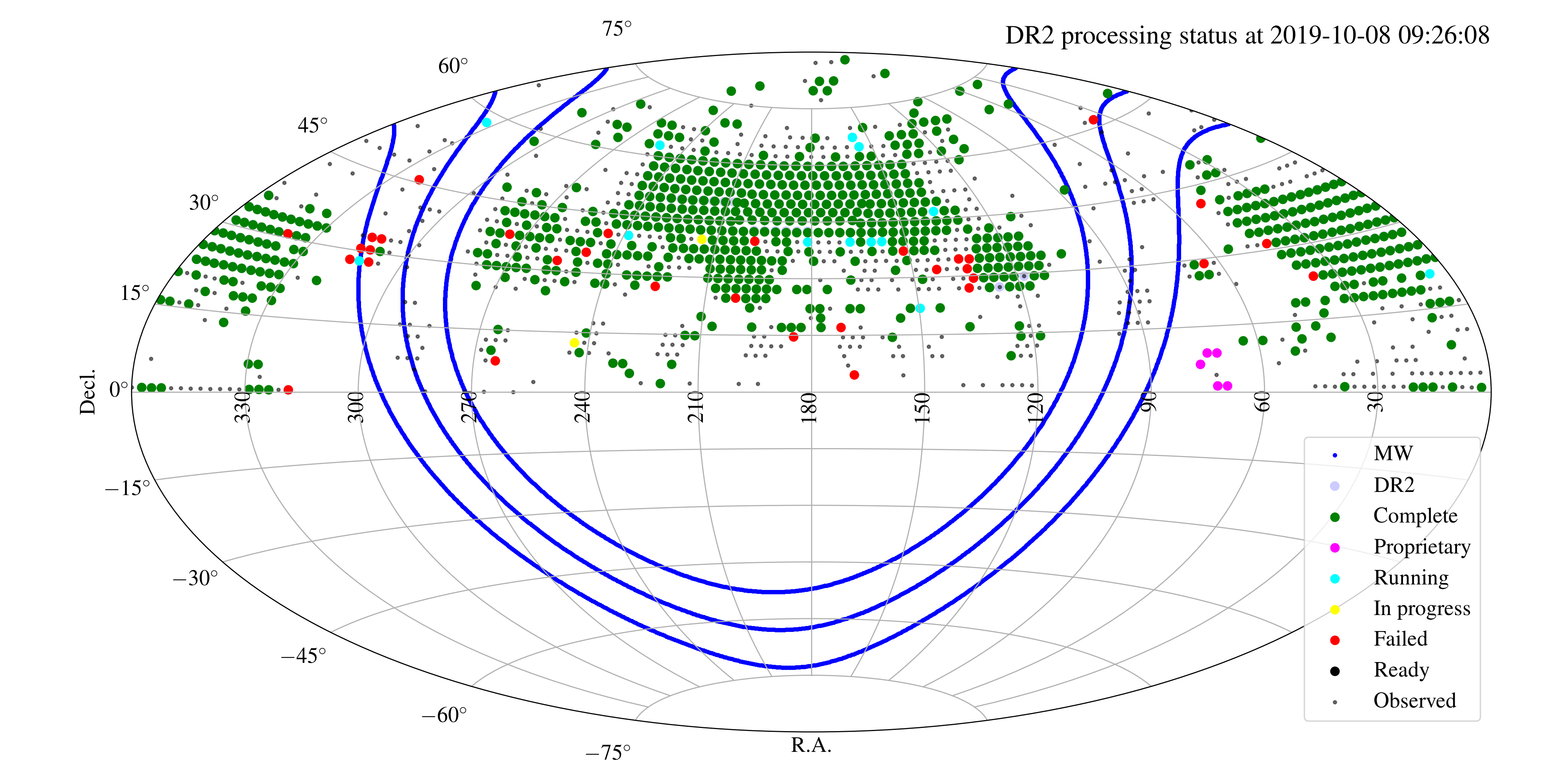}
\caption{Status of the LoTSS as of 8$^{\text{th}}$ Octobre 2019. Green dots mark pointings that are already observed and processed\cite{LoTSS-status}.}
\label{fig:LoTSS_DR2}
\end{figure}

The first data release was published in early 2019\cite{2019A&A...622A...1S} and covers the HETDEX field, which has a size of 424\;deg$^2$, i.e. 2\,\% of the entire survey, see Fig.\,\ref{fig:LoTSS_DR1}.
The source catalogue derived from the first data release contains 325\,694 sources with 231\,716 optical and/or infrared identifications\cite{2019A&A...622A...2W}. A special issue of the scientific journal Astronomy \& Astrophysics (A\&A) has been dedicated to the first twenty-six research papers describing the survey and its first results\cite{specialissue}.
Discoveries made in the framework of the LoTSS cover a multitude of research topics, e.g. detailed studies of black holes in Active Galactic Nuclei (AGN), diffuse radio emission in clusters of galaxies, and cosmic-rays in nearby galaxies, to mention a few (see also Fig.\,\ref{fig:LoTSS_results}).
The second data release of LoTSS will offer the largest source catalogue at radio frequencies to date.
The current status of the processing is shown in Fig.\,\ref{fig:LoTSS_DR2}.

The high sensitivity, resolution and survey speed of LOFAR comes with a price: the large bandwidths, the large distance of stations and the large field of view require entirely novel calibration techniques. LOFAR operates at low frequencies and thus suffers severely from ionospheric disturbances, which can cause decorrelation of the signal and thus limits the sensitivity of the instrument. These challenges are met with non-standard and computational intense methods like advanced directional-dependent calibration techniques\cite{2014arXiv1410.8706T,2016ApJS..223....2V,2018A&A...611A..87T}. Moreover, averaging in time and frequency can only be performed moderately to permit recovering the sky brightness distribution over the entire field of view.
This ensemble of large amounts of data and complex algorithms does not only demand big and robust data archives, but also an efficient and automated data handling in order to achieve high-fidelity radio maps within a reasonable amount of time and computational effort.

\section{The J\"ulich LTA}
The German institutes participating in LOFAR have formed the {\it German Long Wavelength Consortium}\cite{GLOW} (GLOW), which coordinates the German LOFAR activities.
Besides the archival sides in Amsterdam and Pozna\'n, the J\"ulich Supercomputing Centre (JSC) at the Forschungszentrum  J\"ulich (FZJ) operates a large part of the LOFAR {\it Long Term Archive} (LTA). About 50\,\% of all LoTSS observations are archived in the J\"ulich tape archive, i.e., 800 observations occupying about 12\,PByte of storage space on tape.
Such an amount of data needs to be processed in an automated, well reproducible and organised fashion within a reasonable amount of time. An efficient processing therefore demands computing facilities with a fast connection to the data storage in order to avoid unreasonably long transfer times of the data to local computing resources via the internet. Thus, the storing of a large amount of data in the J\"ulich LOFAR LTA and the processing of the data at JSC supercomputers is a significant German contribution to the ILT.
The data in the LOFAR LTAs are among the largest collections of astronomical data.
Operating  these  LTAs  allows  facilities to  explore  strategies  how  to  handle  the  vast  amounts  of scientific  data  generated  by  radio  interferometers. In particular, the staging and the retrieval of the data from tape to the disk pool is a time critical step for the whole processing.

\section{The Automated Processing Scheme for JUWELS}

In the framework of LoTSS the LOFAR Surveys Key Science Project (SKSP) developed an automized job management and execution system for the grid-computing facilities at SURFsara in Amsterdam. These tools are described in Mechev et al. (2017)\cite{2017isgc.confE...2M} and Mechev et al. (2018ab)\cite{2018arXiv180810735M,2018A&C....24..117M}. Since the biggest fraction of LOFAR data observed in the framework of the LoTSS is stored in the J\"ulich LTA, an efficient and fast processing of the data is crucial. A regular 8\,hour pointed observation using the full bandwidth of the telescope at a time resolution of 4\,seconds and a frequency resolution of 48{.}82\,kHz requires about $\sim$8\,-\,16\,TB of disk space, depending on the level of compression. Transferring such an amount of data from the J\"ulich LTA to the computing facilities in Amsterdam via publich network or internet is slow and thus not feasible.

\begin{figure}[htbp]
\centering
\includegraphics[width = \textwidth]{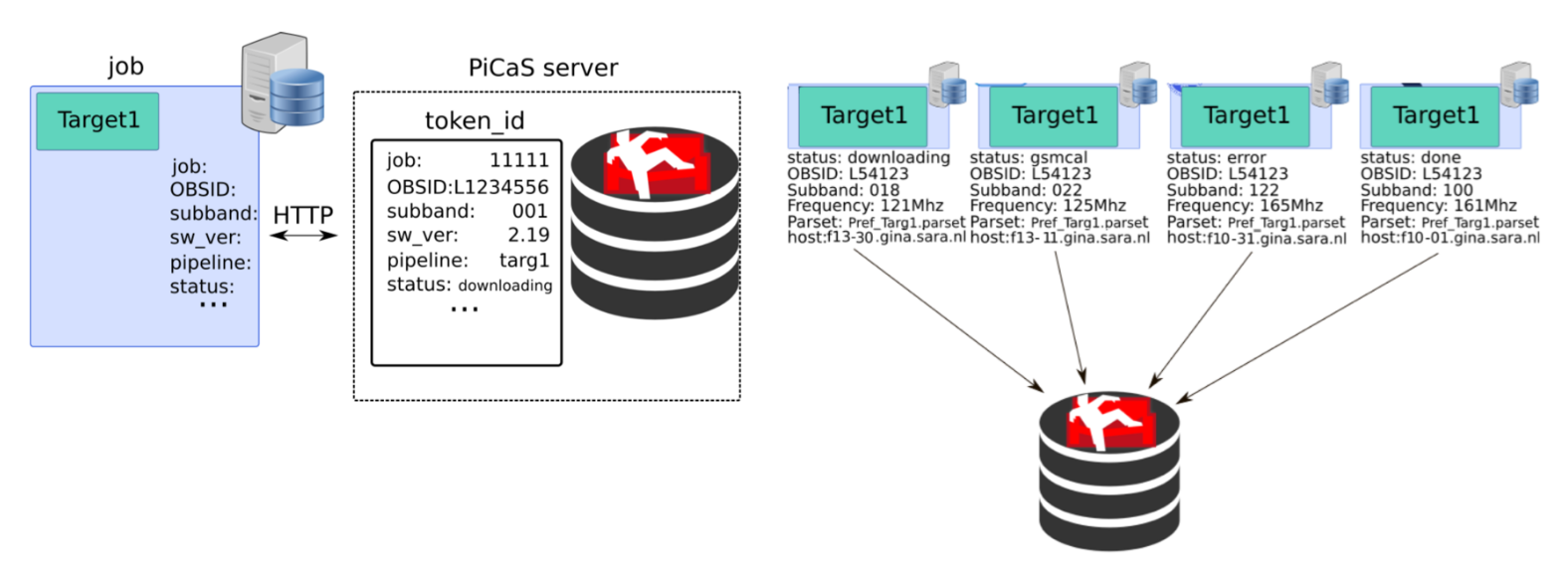}
\caption{Sketch of the internal structure of the job management system developed by the SURFsara group in Amsterdam. Jobs are organised in tokens, comprising of the job description and their corresponding parameters to be used during the processing. All tokens are stored in a database, using Apache couchdb, which can be accessed via HTTP. Image by courtesy of A. Danezi\cite{doc-grid}.}
\label{fig:couchdb}
\end{figure}

The fast disk cache of the J\"ulich LTA is directly connected to the data management server of the FZJ (J\"ulich Data Access Server, JUDAC) and thus provides an order of magnitude higher throughput of data than data transfer via internet. Therefore, we extended the already established automated and coherent processing system at the SURFsara grid-computing facilities in Amsterdam to the supercomputing system JUWELS (J\"ulich Wizard for European Leadership Science). To achieve this goal we developed cluster-specific tools that allow us to retrieve data from the J\"ulich LTA and subsequently process them directly on JUWELS. These tools comprise of:
\begin{itemize}
    \item \textbf{Software installation:} We share a pre-compiled LOFAR installation hosted in a \texttt{cvmfs}-directory at SURFsara. This allows us to perform regular updates via a simple synchronization of the software. The software (parrot-connector) needed to be compiled in user space. Furthermore, cluster-specific environment parameters are adjusted in the retrieved software copy.
    \item \textbf{Job management and execution:}  Service monitoring scripts, running continuously on JUDAC and the head node of JUWELS, act as an interface between the J\"ulich LTA, an external database, and the supercomputer JUWELS, respectively.
\end{itemize}
The external database uses Apache's \texttt{couchdb} and is operated by SURFsara. It serves as an ``order book'' (a so-called ``pilot job framework'') in order to define and manage all the planned jobs, see Figure\,\ref{fig:couchdb}.
All the jobs are organised in so-called ``tokens'' that are comprised of a job description (usually a pre-defined list of tasks or pipelines) and their corresponding parameters (e.g., location of input/output data, observation IDs, software version to be used, current processing status of the job). Thanks to such a central database, from the user's perspective it is not relevant where a LOFAR observation to be processed is actually located and on which or what kind of machine it will be processed. The user does not even need to have any knowledge about the structure of the archive or the supercomputing system to run a job. Moreover, the database allows the user to track the status of the current processing and even inspect all diagnostics and logfiles provided by already finished tasks. Thus, our system is a realisation of the ``near-data processing'' principle.

\begin{figure}[htbp]
\centering
\includegraphics[width = \textwidth]{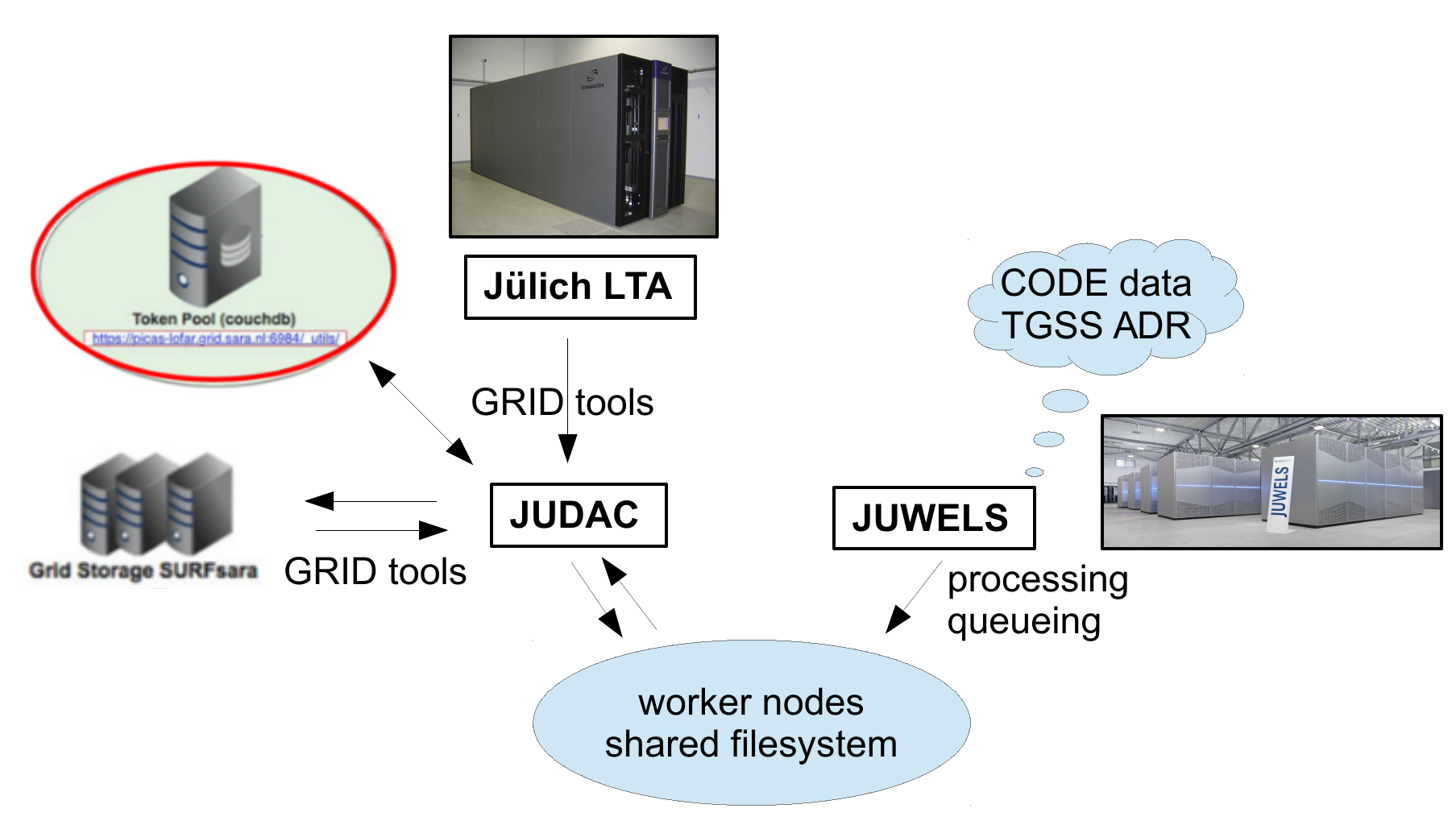}
\caption{Sketch of the automated processing scheme implemented on JUDAC and JUWELS. Interactions that require access to the grid and the external database (here called as Token Pool) are realised on JUDAC. The head node of JUWELS prepares the tasks provided by JUDAC and submits it to the job queue of JUWELS. Before it needs to retrieve online data necessary for the calibration (here denoted as CODE data, a GPS satellite, and TGSS ADR, a skymodel database). Then it awaits execution. The communication between JUDAC and the JUWELS head node is done via the shared filesystem.}
\label{fig:pipeline_scheme}
\end{figure}

In Fig.\,\ref{fig:pipeline_scheme} we present the structure of the processing pipeline. Thanks to the ability of JUDAC to access non-standard ports via HTTP and the availability of grid-tools like \texttt{voms}, \texttt{srmls}, or \texttt{globus-copy-url}, the monitoring service script running on JUDAC takes over the entire communication with the external database and the data retrieval with the J\"ulich LTA. It essentially incorporates the following services:
\begin{itemize}
  	\item[(1)]
  	 interaction with the external database, i.e. looking for new tokens
  	\item[(2)]
  	   checking whether data of a new or active job are staged and transferring staged data from the LTA in J\"ulich to the shared file system, which is accessible for both JUDAC and JUWELS
  	\item[(3)]
  	   preparing environment for the job and writing a so-called ``submission script'' to a dedicated location of the shared filesystem
  	\item[(4)] 
  	  monitoring of submitted jobs via the logfile of an active job, and if applicable reporting back success or failure to the external database
  	\item[(5)]
  	  uploading diagnostic plots created during the calibration of the data and the logfile of the active job to the external database
  	\item[(6)]
  	  transferring the output data to a grid-storage at SURFsara in Amsterdam if job was run successfully
\end{itemize}

A second monitoring script is in place on the head node of JUWELS. It continuously checks for a ``submission script'' put in a dedicated location of the shared filesystem and executes it. Since the worker nodes of JUWELS do not have any internet access, data necessary for the calibration that is only available from online services have to be downloaded before the actual submission of the calibration task to the job queue. So the ``submission script'' contains:
\begin{itemize}
    \item[(1)]
     downloading of all necessary data needed for the particular calibration tasks, e.g. skymodels from an online repository
    \item[(2)]
     submitting the job to the JUWELS queue
\end{itemize}
 After submission the ``submission script'' is deleted and the job will be only traced by the monitoring script on JUDAC via the job ID returned from the JUWELS queueing system, in particular if in execution via the written logfile.

After processing the reduced data and all derived solution tables are transferred to a central grid-network storage hosted by SURFsara where it is accessible for all members of the consortium for further advanced processing.
The code and its documentation are available on github\cite{github}.

\section{LOFAR Data Reduction in J\"ulich: prefactor}
The \texttt{prefactor}\cite{prefactor,2019A&A...622A...5D} pipeline is a core component of the LOFAR software and performs the first direction-independent calibration, which is of fundamental importance for the success of all upcoming direction-dependent calibration and image reconstruction steps.
An average LoTSS observation with a duration of 8\,hours is bookended with calibrator observations with a duration of ten minutes. The processing of the data on JUWELS with \texttt{prefactor} is therefore split into two main parts:
    \begin{itemize}
  	\item[(a)]
  	   {\bf calibrator pipeline:} extraction of instrumental effects from the calibration of the calibrator observation
  	\item[(b)]
  	   {\bf target pipeline:} correcting for the instrumental effects (derived from the calibrator pipeline) in the actual target data, flagging, further averaging of the target data and further calibration off a model of the field
  \end{itemize}
In case of (a) it derives corrections for instrumental effects like the polarisation alignment, the bandpass, and the clock offsets (mainly for remote stations). To achieve this the instrumental effects have to be separated from the ionospheric effects, i.e., phase signal delays and Faraday Rotation of the phase signals due to the time-dependent variation of the total electron content (TEC) at the location of the LOFAR station and its pointing direction. \texttt{prefactor} calibrates and corrects for all these effects in the order of their interference with the signal along its path from the sky to the telescope and even uses the full bandwidth if necessary to increase the signal-to-noise ratio. 

In the target pipeline (b) the derived corrections from the calibrator pipeline are applied to the target data. These corrections allow for moderate averaging and thus reducing the data size without decorrelating the signal. Moreover, in the course of the averaging the international stations are discarded. In a second step an initial direction-independent phase-only self-calibration with a skymodel is performed to reduce still remaining residuals in the phase signal. The reduction in data size achieved is usually about a factor of 64 and thus allows for a much easier data transfer and handling in the following direction-dependent calibration steps. Future full-resolution surveys, making use of the international stations, will be more challenging, since they will not allow for such high levels of averaging.

\texttt{prefactor} offers many diagnosis tools to assess the quality of the calibration in a purposeful way. Dynamic spectra plots of the calibration solutions of each individual antenna for every calibration step allows the user to easily spot failures in the processing. Missing data (of the calibrator or target) or data of low quality are identified, e.g. through flagging of stations that are severely affected by Radio Frequency Interference (RFI), or large frequency regimes that are affected by broad-band RFI. This mitigates the need for repeating the processing with adapted parameters and will lead to a successful calibration run of a random observation in more than 90\% of the cases.

\section{Concluding Remarks}
Efficient and robust processing of large amounts of scientific data is a key challenge for the future of radio interferometry. Large-scale surveys like LoTSS allow for deep insights into previously unveiled regimes of the radio sky. With about 30\,PB of required storage space, LoTSS will be among the biggest scientific data collections in the world. We implemented a set of service scripts for the supercomputing system JUWELS that have so far enabled us to retrieve and calibrate the data of 500 LoTSS observations stored in the LTA in J\"ulich in an automated, robust and efficient manner with prefactor. Hereby, we make use of an already established framework hosted and maintained at SURFsara in Amsterdam, which was originally developed for processing LoTSS data at their grid-computing facilities. This allows an external user to manage the jobs for processing the data with a single interface independent on the used hardware or software infrastructure and is thus a realisation of the principle of ``near-data processing''.
The ILT has identified the need to develop all three LOFAR LTAs, including the one operated by the JSC, into integrated LOFAR Science Data Centres, which would combine archiving and computing. The Science Data Centres would offer standard pipelines to all LOFAR users and coordinate the pipeline development. The work done by the SKSP at SURFsara and JSC represents a first step towards achieving this goal.

\section*{Acknowledgments}
AD acknowledges support by the BMBF Verbundforschung under the grant 05A17STA.
JBRO acknowledges financial support from NWO Top LOFAR-CRRL project, project No. 614.001.351.
The authors gratefully acknowledge the Gauss Centre for Supercomputing e.V. (www.gauss-centre.eu) for funding this project by providing computing time through the John von Neumann Institute for Computing (NIC) on the GCS Supercomputer JUWELS at J\"ulich Supercomputing Centre (JSC).
This work was carried out on the Dutch national e-infrastructure with the support of the SURF Cooperative through grants e-infra 160022 \& 160152. The authors would like to thank the staff at SURFsara for all assistance received during this project.
LOFAR (van Haarlem et al. 2013) is the LOw Frequency ARray designed and constructed by ASTRON. It has observing, data processing, and data storage facilities in several countries, which are owned by various parties (each with their own funding sources) and are collectively operated by the ILT foundation under a joint scientific policy. The ILT resources have benefited from the following recent major funding sources: CNRS-INSU, Observatoire de Paris and Université d’Orléans, France; BMBF, FZ J\"ulich, MIWF-NRW, MPG, Germany; Department of Business, Enterprise and Innovation (DBEI), Ireland; NWO, The Netherlands; The Science and Technology Facilities Council (STFC), UK.

\end{document}